\documentstyle[12pt,cite]{article}                                        
\catcode`\@=11                                                       
\def\ee{\end{equation}}
\def\be{\begin{equation}}

\def\H{{\cal H}}

\def\1{{\bf 1}}

\def\A{{\cal A}}
\def\F{{\cal F}}
\def\s{{\sc s}}
\def\Z2{{\bf Z}_2}
\def\H{{\cal H}}

\input epsf
\newcounter{fignum}
\newcommand{\figuurnum}{\arabic{fignum}}

\newcommand{\figuurplus}[3]{
\addtocounter{fignum}{1}
\addcontentsline{lof}{figure}{\protect
\numberline{\arabic{section}.\arabic{fignum}}{#3}}
\hspace{-3mm}{\it Fig.}\ \figuurnum.
\begin{figure}[t]\begin{center}
\leavevmode\hbox{\epsfxsize=#2 \epsffile{#1.eps}}\\[5mm]
\parbox{10cm}{\small \bf Fig.\ \figuurnum: \it #3}
\end{center} \end{figure}\hspace{-1.5mm}}

\begin{document}                                                     

\renewcommand{\theequation}{\thesection.\arabic{equation}}        
\newcommand{\mysection}[1]{\setcounter{equation}{0}\section{#1}}  

\begin{titlepage}

\hfill{SPIN 2000/14}

\vspace{1 cm}

\centerline{{\huge{
Six Dimensional Topological Gravity and }}}
\centerline{{\huge{
the Cosmological Constant Problem
}}}

\vspace{2 cm}

\centerline{\large{G. Bonelli and A.M. Boyarsky}}
\vspace{1 cm}
\centerline{Spinoza Institute, University of Utrecht}
\centerline{Leuvenlaan 4, 3584 CE Utrecht, The Netherlands}
\centerline{bonelli@phys.uu.nl $\quad$ boyarsky@phys.uu.nl}

\vspace{4 cm}

{\it Abstract}:
We formulate a topological theory in six dimensions with gauge group
$SO(3,3)$ which reduces to gravity on a four dimensional defect if suitable 
boundary conditions are chosen.
In such a framework we implement the reflection automorphism
of $SO(3,3)$ as a $\Z2$ symmetry
which forbids the appearance of a gravitational cosmological constant.
Some temptative speculations are presented also for the possible inclusion of 
the matter contribution at a full quantum level.

\vspace{3 cm}

PACS: 98.80.E ; 04.50

\end{titlepage}

\section{Introduction}

The fine tuning of the cosmological constant is one of the open problems in 
modern theoretical physics \cite{w}. In few words, the Hilbert-Einstein action
$S_{HE}=\frac{1}{16\pi G}\int R\sqrt{g}d^4x$
admits a natural extension to
$S_{HE}+\Lambda \int\sqrt{g}d^4x$
for any constant real $\Lambda$. From the gravitational point of view 
the natural scale for $\Lambda$ is the gravitational 
inverse volume factor $\frac{1}{G^2}\sim M_{Plank}^4$.

On another side, a sensible non zero additive contribution to the 
cosmological constant is expected on general grounds from the quantum theory
of the Standard model and to be of the order $M_{SM}^4$ as a vacuum energy.

Actually none of these two scales are observed.
The present experimental data \cite{cc} in fact 
predict $\Lambda$ to be absolutely negligible with respect 
to both the natural gravitational inverse volume 
and the Standard model one.

The fine tuning of the cosmological constant represents then a problem
since it seems unnatural that two contributions depending on two
independent scales do cancel a priori.

A usual mechanism in field theory to tune to zero a parameter
is to impose a symmetry which prohibits it. A promising candidate for this
role is supersymmetry which naturally demands zero vacuum energy.
One obvious difficulty with supersymmetry from this prospective
is that it should be broken at the Standard Model scale while the 
cosmological constant is still negligibly small.

On another side, in recent times, a phenomenological new
framework arose in considering the possible pros in having 
extra dimensions \cite{extra} where gravitational degrees of freedom 
can propagate. 
In this letter we will begin our analysis by considering a different type 
of extra dimensions. 

Let us describe our starting point by recalling a well known fact.
Any theoretical physicist lives with its own signature for the
space-time metric: depending on taste 
one can choose either $(-,+,+,+)$ or $(+,-,-,-)$.
Both possibilities are equivalent 
with respect to $S_{HE}$
up to a global sign reversal, but they differ 
if the cosmological term is added. The latter does not take any sign
reversal and therefore one relates the two conventions by 
$$
S_{HE}+\Lambda\int \sqrt{g}
\quad
\to
\quad
-\left(S_{HE}-\Lambda\int \sqrt{g}\right)\,.
$$
In this letter we will promote this convention choice to a $\Z2$
symmetry which in principle could prohibit the cosmological term.

This could be achieved by embedding the relative 
local $SO(1,3)$ Lorentz group into a larger $SO(3,3)$
gauge group governing a six dimensional gravitational field theory.
In such an enlarged theory one can consider inequivalent embeddings of 
$SO(1,3)$ in $SO(3,3)$. These embeddings are in two classes 
which are exchanged by the signature changing operator $\s$
\footnote{
As we will see, a very important point is represented by the fact that $\s$ 
is an outer automorphism of $SO(3,3)$ in the sense that $\s$ itself
is not a $SO(3,3)$ group element. This is explained in more details in 
the Appendix.}.

A first difficulty that one encounters in trying to perform such a 
construction is that, if the six dimensional field theory admits 
propagating degrees of freedom, from the point of view of an embedded 
four dimensional world the appearance of tachionic transverse
excitations takes place.
Therefore, to avoid this unwelcome circumstance, 
the most obvious possibility is to restrict the class 
of possible six dimensional theories to that of 
topological field theories in such a way that 
any propagation of degrees of freedom in its 
bulk space-time does not take place at all \cite{'tHooft}.
Let us here note that the appearance 
of a possible role played by topological field theories
in a framework which could cancel the gravitational cosmological
constant was already advocated in \cite{lpw}.
In this letter we will build a minimal model which fulfills all the above 
properties.

After it became clear that three dimensional 
gravity can be reformulated as a Chern-Simons theory \cite{ATW}
there has been a lot of attempts to understand if four dimensional gravity
has something to do with some other topological theory.
Here we present a construction of classical four dimensional gravity as 
a six dimensional topological theory which induces dynamical degrees of 
freedom on a four dimensional defect if properly coupled to it.

Then, using this result,
we will describe a mathematical realization of
a discrete $\Z2$ symmetry which prohibits the appearance of a non zero 
cosmological constant in classical gravity theory.
We will finally implement such a symmetry quantum mechanically.

Let us take henceforth a minimal perspective in which we try to realize 
such an higher theory as a topological gauge theory in six dimensions 
with gauge group $SO(3,3)$ and just a $so(3,3)$ valued connection
$\A$ as a fundamental field. In other words, we consider a topological 
gravity theory in six dimensions consisting only of a spin connection and 
not of a couple spin connection and sechs-bein.

A second difficulty that one meets in this model is 
related to the implementation of the $\Z2$ symmetry.
The group $SO(3,3)$ admits a $\Z2$ automorphism $\s$ which
exchanges $\eta\,\to\,\eta^\s=-\eta$ and acts as an outer automorphism
on the group. Its invariant subgroup is isomorphic to $GL(3)$.
Therefore, if one tries to restrict naively any $so(3,3)$ six dimensional
connection in an $\s$-invariant way, one can at most obtain a 
four dimensional $GL(3)$ gauge group which, unfortunately, 
can not contain any $SO(1,3)$ gauge subgroup.
This fact implies that the simplest embedding procedure has to implement
the signature changing operator $\s$ on a couple of $\s$-conjugate 
configurations $\A^{(\sigma)}$, with $\sigma=\pm1$, 
whose content will be properly fixed afterwards
to obtain a symmetric configuration
\footnote{
Let us note that an apparently different way could be tried by 
embedding the problem in a complexified gauge group as
$SO(3,3)\,\to\, SO(6,{\bf C})$ where the signature changing operator
would act as an inner automorphism singling out a fixed complex structure
in $SO(6,{\bf C})$. This again naturally leads 
to a similar doubling of the degrees of freedom.}.

The main idea that we will work out here is the following. 
We consider a topological field theory in a six dimensional space $Y$ with 
boundary such that no dynamics is induced on the local degrees of freedom 
in the interior of $Y$. Then non trivial dynamics 
can only take place at the boundary. 
We couple the boundary to the 
bulk fields by specifying their boundary conditions.
What remains is then a dynamical theory on the boundary for the unfixed part
of restricted bulk fields.
We can generate such a situation in a suitable regularized version of
$Y\setminus W$, where $W$ is a four dimensional submanifold of $Y$.
In this case the boundary theory can be then resolved to a four dimensional 
theory on $W$.

In the following we will first give a mathematical account for 
the situation described above and then, in a subsequent section,
we will explain how to use this mathematical construction for 
classical gravity and how to include in the model any quantum matter 
contribution.

\section{From six to four dimensions}

Let $Y$ be a (compact) six dimensional manifold and $W$ a four dimensional
sub-manifold embedded into $Y$. 
Let $\tau$ be the Poincare' dual of $W$, i.e.
a closed 2-form on $Y$ such that $\int_Y \tau\wedge O^{(4)}=
\int_W O^{(4)}|_W$ for all 4-forms \footnote{
Locally it just means that if, e.g., $W$ is embedded like $x_1=x_2=0$ 
(for some coordinates $x_1 \ldots x_6$ ), we can choose 
$\tau =dx_1dx_2\delta(x_1)\delta(x_2)$.} in $Y$.
Really $\tau$ is defined up to a full differential and so, being a cohomology 
class, we can choose various representatives for it.
Following \cite{bt} we can choose 
$\tau=d\left(\rho\frac{d\theta}{2\pi}\right)$,
where $\rho$ is a bump function, which is equal to zero on $W$ and to $-1$
far enough from it while $\theta$ is a local angular coordinate along the 
transverse $S^1$.
Let $\A_{AB}=A^\mu_{AB}dx_\mu$ be a connection one form for a given $SO(3,3)$ 
bundle ${\cal X}$ on $Y$ and $\F_{AB}=d\A_{AB}+\A_{AC}\wedge\A^C_B$ be its 
curvature two form valued in $so(3,3)$.

If $Y$ is compact, the functional 
$I_6=c\int_Y\epsilon^{ABCDEF}\F_{AB}\wedge\F_{CD}\wedge\F_{EF}
=c\int_Y \F^3$
is independent on the particular connection
and defines an invariant of the bundle. 
Under a properly chosen $c$ prefactor, $I_6$ is, in this case, 
integer valued.

If $Y$ has instead a boundary, this is no more true. In this case
$I_6$ depends upon the value of the connection on the boundary
of $Y$.
To see this, for example, notice that if the bundle is trivial we can rewrite
$I_6=\int_{\partial Y} [C-S]$ as the integral of 
the corresponding Chern-Simons form on the boundary.

To adopt the above construction to the case of our interest, we have to make 
some more preliminary steps.
Fix an auxiliary euclidean metric $\gamma$ in $Y$ and consider the 
space of points in $Y$ which have $\gamma$-distance less than
a given $\varepsilon>0$ from $W$. Call this space $B_{\varepsilon}^W$
and $Y_\varepsilon=Y\setminus B_{\varepsilon}^W$ its complement in $Y$.
$Y_\varepsilon$ is an $\varepsilon$-regularization for $Y\setminus W$.
Notice that the disk $\partial Y_\varepsilon$ is a realization of the 
total space of the normal bundle of $W$ in $Y$ \cite{bt}. 
This means that locally $\partial Y_\varepsilon$ is like $W\times S^1$.
From now on the limit $\varepsilon\to0$ will be always implied
in our formulas.

Now, to specify a coupling of the four dimensional defect to the bulk 
theory in this smoothed picture, we can choose
some boundary conditions for $\A_{AB}$ at
$\partial Y_\varepsilon$.
For this, let's make the following index splitting along
an $SO(2)\times SO(1,3)$ subgroup with $x,y,\dots=1,2$ $SO(2)$ indices 
and $a,b,\dots=0,1,2,3$ $SO(1,3)$ indices. Explicitly we define
$\A_1^2=\omega$, $\A_{a1}=e_a$, $\A_{a2}=\pi_a$
and $\A_{ab}=\omega_{ab}$. Let us remember that all these objects are 
1-forms and so $\omega_{ab}=\omega_{ab}^\mu dx_\mu$ and so on
\footnote{
Let us also notice here that this splitting make sense only locally if 
the bundle ${\cal X}$ at $W$ is irreducible. If it is instead reducible
as ${\cal X}|_W={\cal U}\oplus {\cal H}$, with ${\cal U}$ an $SO(2)$
bundle and ${\cal H}$ an $SO(1,3)$ bundle to be identified with
the tangent bundle $T(W)$, then
$\omega_{ab}$ is a connection for ${\cal H}$ and a singlet for ${\cal U}$,
$(e_a,\pi_a)$ are two frames for ${\cal H}$ and a doublet for ${\cal U}$
and $\omega$ is a singlet for ${\cal H}$ and a connection for ${\cal U}$.}.

Under these definitions, we get for the gauge curvature elements the 
following expressions
$$
\F_{12}=\sigma d\omega -e^a\wedge\pi_a
$$
$$
\F_{a1}=De_a+\omega\wedge\pi_a
$$
$$
\F_{a2}=D\pi_a-\omega\wedge e_a
$$
$$
\F_{ab}=R_{ab}-\sigma (e_a\wedge e_b+\pi_a\wedge\pi_b)
$$
where $\sigma=\pm 1$ is fixed by $\A_{12}=\sigma\omega$
and corresponds to the signature of the $SO(2)$
within $SO(3,3)$, $De_a=de_a+e_b\wedge\omega^b_a$,
$D\pi_a=d\pi_a+\pi_b\wedge\omega^b_a$ and $R_{ab}=d\omega_{ab}+
\omega_a^c\wedge\omega_{cb}$.

We specify our boundary conditions for the connection as the identification 
of the $SO(2)$ sub-bundle with the normal bundle of the defect ${\cal N}(W)$,
which is ${\cal U}|_W=\sigma{\cal N}(W)$,
as $d\omega=\tau$
\footnote{Such that the freedom of choosing a particular 
$\omega$-connection representative becomes the choice of a potential 
for $\tau$, while the $SO(2)$ gauge invariance is the usual invariance of 
$\tau$ under a shift of an exact differential of the potential}
and fixing also one half of the transverse
mixed connection components as $\pi_a|_W=0$.
After this boundary conditions are imposed, the degrees of freedom 
left on the defect are $(e_a,\omega_{ab})$ which are valued in 
an $iso(1,3)$ subalgebra of our initial $so(3,3)$.

Now, we can calculate the actual value of the $I_6$ functional.
Under the above decomposition we have
$$
\F^3=6\left(\F_{12}\epsilon^{abcd}\F_{ab}\F_{cd}-
4\epsilon^{abcd}\F_{ab}\F_{c1}\F_{d2}\right)\, .
$$
From this we recognize immediately that $\F^3$ is at most linear
in $\omega$ and in $d\omega$ with generically a further 
inhomogeneous term. 
Let us recall that $\F^3$ is always 
\footnote{If the bundle is non trivial this is of course 
just a local statement, in the sense that the Chern-Simons form
is not a well defined covariant form. 
In this section we will assume the bundle 
to be trivial, postponing the discussion about the non trivial bundle case 
to next section.}
the total differential of the relative Chern-Simons form
and in our case $\F^3=d(\omega\wedge X^{(4)} +Z^{(5)})$, 
where $X^{(4)}$ and $Z^{(5)}$ are respectively a 4-form and a 5-form
which are independent on $\omega$.
If the bundle is reducible in the sense above, 
the actual form of $I_6$ 
can be exactly calculated by Stoke's theorem as
$$
I_6|_{b.c.}=c\int_{\partial Y_\varepsilon}\omega\wedge X^{(4)} + Z^{(5)}\, ,
$$
where, with our boundary conditions,  
$$X^{(4)}=\sigma\epsilon^{abcd}
\left[R_{ab}-\sigma (e_ae_b)\right] 
\left[R_{cd}-\sigma (e_ce_d)\right]$$ and $Z^{(5)}$ vanishes.
Choosing a representative for the $W$ cohomology element
and performing the $\varepsilon\,\to\,0$ limit
we can easily reduce this integral to a four dimensional one. 
Solve, in fact, $d\omega=\tau$
as $\omega=\rho \frac{d\theta}{2\pi}+d\phi$, for some scalar function $\phi$,
and perform the integral along the transverse circle
\footnote{To get $\int_{\partial Y_\varepsilon}d\phi\wedge X^{(4)}=0$ 
one has really to use also the continuity properties at $W$ of the 
involved fields.}.
In formulas we get that
$$
I_6|_{b.c.}=-c\int_W X^{(4)}=
-c\left[\sigma\int_W \epsilon^{abcd}R_{ab}\wedge R_{cd}
\right.$$ $$ \left.
-2 \int_W \epsilon^{abcd}R_{ab}\wedge e_c\wedge e_d
+\sigma\int_W \epsilon^{abcd}e_{a}\wedge e_b\wedge e_c\wedge e_d
\right]\, .
$$
The result is then the H-E action on $W$ augmented by 
a cosmological term and a term proportional to the
Euler characteristic of $W$.
Notice that the H-E action does not depend on the signature
$\sigma$ of the embedded $SO(2)$, while the other two terms do.

\section{The symmetric setting}

The construction that we considered in the last section is not symmetric
under the operation of changing the
signature of the embedding of the normal bundle. 
One possible way to have a symmetric situation is to double the gauge bundle.
This way we are able now to couple our defect symmetrically 
with respect to $\s$-conjugation.

Let us therefore introduce two $SO(3,3)$ bundles ${\cal X}_\sigma$,
labeled by $\sigma=\pm 1$, on 
$Y$ and their corresponding connection one forms $\A^{(\sigma)}_{AB}$.
Then we consider as a total action the sum of the two $\int_{Y_\varepsilon}
\left(\F^{(\sigma)}\right)^3$ terms under the following boundary conditions.
We assume the two bundles to be reducible to ${\cal U}_\sigma\oplus{\cal H}$
where ${\cal H}$ is a common $SO(1,3)$ factor and ${\cal U}_\sigma$
are two $SO(2)$ bundles such that
${\cal U}_\sigma|_W=\sigma{\cal N}(W)$.
We set also $\pi^{(\sigma)}_a=0$ on $W$
together with $e_a^{(+1)}=e_a^{(-1)}=e_a$ and 
$\omega_{ab}^{(+1)}=\omega_{ab}^{(-1)}=\omega_{ab}$.
Therefore, on the boundary, $\F_{12}^{(\sigma)}=\sigma\tau$
and $d\omega^{(\sigma)}=\tau$.

We then obtain the following total action
$$
I_{tot}|_{b.c.}=I_6|_{b.c.}^{(-1)}+I_6|_{b.c.}^{(+1)}=
-\sum_{\sigma=\pm 1}
c\left[\sigma\int_W \epsilon^{abcd}R_{ab}\wedge R_{cd}
\right.$$ $$\left.
-2 \int_W \epsilon^{abcd}R_{ab}\wedge e_c\wedge e_d
+\sigma\int_W \epsilon^{abcd}e_{a}\wedge e_b\wedge e_c\wedge e_d
\right]$$ $$
=
4c\int_W \epsilon^{abcd}R_{ab}\wedge e_c\wedge e_d
$$
which is the H-E action without any cosmological term
\footnote{
In Section 2 we left aside the possibility that the six dimensional 
bundles are non trivial.
In such a case we can not use Stoke's theorem directly on the action.
Nonetheless, under infinitesimal variations of the connection,
$\int_{Y_\varepsilon}\F^3$
reacts only to the change of the connection at the boundary. In fact
$\delta\int_{Y_\varepsilon}
\F^3=3\int_{\partial Y_\varepsilon}
\epsilon^{ABCDEF}\delta\A_{AB}\wedge\F_{CD}\wedge\F_{EF}$,
where we used the Bianchi identity and Stoke's theorem.
Under our boundary conditions, which then fix partly the freedom in varying 
the connection on the boundary, and summing over the two $\s$-conjugate 
copies of the connection one finally gets that
$
\delta I_{tot}|_{b.c.}=\delta \left\{
4c\int_W \epsilon^{abcd}R_{ab}\wedge e_c\wedge e_d
\right\}\, .
$
From this we conclude that, since their equations of motion coincide, 
also in the case of non trivial bundles 
our topological theory is equivalent to gravity on $W$ without the 
cosmological term.
Let us stress that in this case the total action $I_{tot}|_{b.c.}$
possibly depends on global degrees of freedom in the bulk
related to the structure of the two reduced $SO(3,3)$ bundles 
${\cal X}^{(\sigma)}$.}.

The result of our construction is that, being able to map 
gravity in four dimension to an equivalent six dimensional topological 
theory, we find a specific $\Z2$ symmetry which differs between
the H-E action and the cosmological term and once it is implemented 
cancels the latter.

\subsection{Inclusion of the matter sector}

Now we want to speculate about a possible way to extend our construction 
to a slightly more realistic situation in which also quantum matter degrees 
of freedom are present on the four dimensional defect.

In the previous section we described a topological gauge theory in six
dimensions which is equivalent to four dimensional gravity.
In such a framework the cosmological term was canceled by
the signature changing symmetry of the theory.

To implement our symmetry, we need a little further step.
We want here to notice a trivial identification of two procedures in
the above construction. The first procedure that we consider is the 
construction that we did in the previous section: we have chosen
one four dimensional defect coupled to both the connections in a 
symmetric way via specific boundary conditions. Let us call this 
procedure I. Another possibility to obtain the same result is 
to consider two four dimensional defects $W_\sigma$, with $\sigma\pm1$
each of them being coupled only to a single connection $\A^{(\sigma)}$
as in Section 2. We then consider them to be ``close enough'' in such a way 
that we are able to identify all degrees of freedom on them. 
But our six dimensional theory
is topological and all non-zero bulk distances are physically equivalent
from its point of view. Therefore the only possibility is that the two defects
actually coincide as $W_\sigma=W$. 
This second construction leads exactly to the same result as
before and we call it procedure II.
The two equivalent procedures are described in
\figuurplus{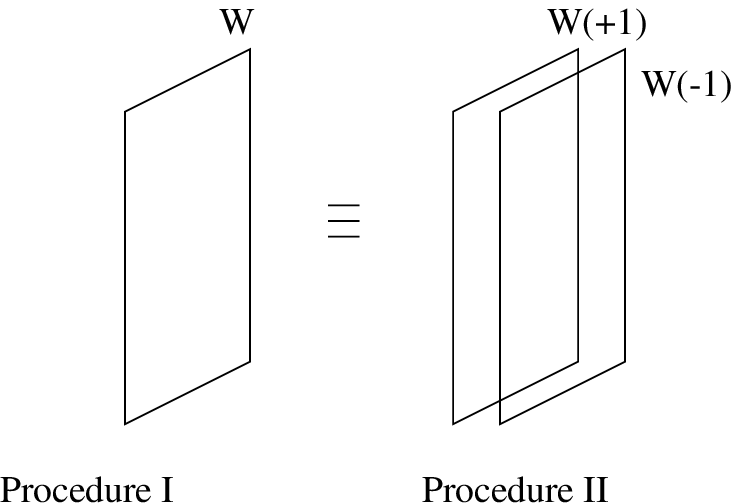}{6 cm}{``Close enough'' in a topological theory means
homotopically equivalent. The two equivalent procedures.}
Let us now implement the equivalence of procedure I and II 
at a quantum level in an hamiltonian formulation.
From the point of view of procedure II our Hilbert space can be considered
to be of the form $$\H_{phys}=\H\otimes\H/{\sim}\, ,$$
where $\sim$ is the following equivalence.
Let any excitation on one of the two factor be created by its 
relative creation operator $a^\dagger$ and destroyed by $a$.
Then, given any state
$|\psi>\otimes |\eta>$, we implement the equivalence with 
procedure I, by representing the excitation
in two conjugate equivalent ways as
$$|\psi>\otimes\left(a^\dagger |\eta>\right)\,\sim\, 
\left(a |\psi>\right)\otimes |\eta>\, .$$
Therefore we can choose in principle a picture to represent 
our physical Hilbert space by fixing a given state in, let's say, the 
second factor as
$\H_{\bar\eta}=\H\otimes |\bar\eta>$. 
Since we are representing as equivalent particles in the first factor and 
anti-particles in the second one, the natural evolution operator is in the form
$U=u\otimes u^\dagger$, where $u={\rm e}^{ith}$ 
is the evolution operator on a single copy of our system.
As a consequence, we find that total hamiltonian
$H=h\otimes1-1\otimes h$ is odd under the 
$\Z2$ permutation of the two factors of the Hilbert space.

Let us now note that the only representations which are stable under
the time evolution are the ones for which $|\bar\eta>$ is an eigenvector 
for the hamiltonian $h$, i.e. $h|\bar\eta>=e_{\bar\eta}|\bar\eta>$.
The vacuum state in a generic stable representation is
represented by $|0>\otimes|\bar\eta>$, where $|0>$ is the vacuum
state for $h$.
Let us now ask for a $\Z2$ symmetric vacuum.
This condition reads as $|0>\otimes|\bar\eta>\,\sim\,|\bar\eta>\otimes |0>$
and is solved only by $|\bar\eta>=|0>$.
We can now write down our total hamiltonian in the only symmetric 
representation $\H_{s}=\H\otimes |0>$ of $\H_{phys}$ as follows
$$H=h\otimes1-1\otimes h=(h-e_0)\otimes1\, ,$$ 
where $e_0$ is the lowest eigenvalue of $h$ and, by definition, $h|0>=e_0|0>$.

We can now calculate the vacuum energy of our system, which is its total 
contribution to the cosmological constant, to be zero as
$$
E_{vac}=(<0|\otimes<0|)H(|0>\otimes|0>)=e_0-e_0=0\, .
$$

\section{Conclusions}

In this letter we addressed the formulation of a new parity symmetry of the 
vacuum which, once it is implemented also at a quantum level, suggests a natural 
mechanism to cancel the cosmological constant. This is done by redrawing 
our world as a four dimensional defect in six dimensions and its 
gravitational degrees of freedom as governed by a topological theory in 
the bulk ambient space.

Let us speculate on the physical relevance, if any, 
of the six dimensional picture.

The obvious disadvantage of our model is that it may seem to be 
too artificial now.
Indeed, one can say that, from four dimensional point
of view, we are just finding a phenomenologically motivated possibility
to prohibit the cosmological constant which is based on the
observation that the latter has wrong transformation property under
signature changing. The first goal of this paper was indeed to show a new
mechanism for a simple symmetry to
single out a vacuum whose energy is naturally zero.
From this point of view,
we can think that the six dimensional construction 
is just a mathematical device 
to render material our picture and that the two extra dimensions 
are metaphysical. 

On the other hand, this symmetry of the vacuum turns out to be invisible to 
our world (i.e. we did not assumed any particular structure for the 
hamiltonian $h$ in the previous section)
and therefore, if these two extra -- dimensions 
are metaphysiscal we do not have any a priori physical reason to insist
on the reflection symmetry. Notice however that, if present, this symmetry
turns out to be naturally stable under possible quantum corrections 
due to the fact that the two mirror sectors do not interact
\footnote{
The four dimensional picture appearing at this point resambles at some extent
the one discussed in \cite{linde}. We would like to thank 
E. Kiritsis and V. Mukhanov for bringing this paper to our attention after 
this paper was completed.
}.

Another possibility is that our picture should be considered as a 
small building 
block which could be included in a deeper ultimate theory.
A minimal way in this direction, from the point of view of 
the present picture, would be to give a six dimensional origin also 
to matter fields. This is beyond the goal of this letter, but we 
want to make one more final observation.
A possible attitude could be to try to take more seriously
the pure topological theory in the bulk and consider the degrees of 
freedom that we already have in the game. 
These are four dimensional defects coupled in the way 
that we described before. Suppose one lives on one of them $W$
and feels the full six dimensional theory from there. 
If there are other dynamical defects $W'$, he will 
have experience of them only trough their intersection with
its own world $W\cap W'$. These intersections are two dimensional and might appear,
in his given four dimensional world, as strings. 
This is due to the following decomposition $T(W)|_{W\cap W'}=T(W\cap W')\oplus 
{\cal N}(W')$ and to the facts that $T(W)$ is a $SO(1,3)$ bundle and ${\cal N}(W')$
a $SO(2)$ one. Therefore the only possible decomposition pattern for $T(W)$ is with
$T(W\cap W')$ being a $SO(1,1)$ bundle.
We see then that, from this point of view, it would seem possible to understand the 
six dimensional topological theory in which also the four dimensional defects
became dynamical as a string theory when experienced from a single given 
defect.

\vspace{.3 cm}
{\bf Acknowledgements}$\quad$
Let us acknowledge very interesting and encouraging
discussions with G. 't Hooft. We want also to thank 
M. Bertolini,
C. Hoffman,
A. Neronov,
and M. Serone
for discussions.

{\bf Appendix: Some useful facts about $SO(3,3)$ and $so(3,3)$}

$SO(3,3)$ is the group
of rotations preserving the form
$\eta={\rm diag}(1,1,1,-1,-1,-1)$
as $R^t\eta R=\eta$ and with unit determinant.
It admits an external automorphism
given by the signature changing operator $\s$
which we identify with the adjoint action of the 
matrix element
$S=\left(\matrix{0 & \1_3 \cr \1_3 & 0 \cr}\right)$
as $\s R=SRS$. In fact, multiplying on the left and on the right
side $R^t\eta R=\eta$ by $S$ and using $S^2=\1_6$ and $S^t=S$, we get
$(\s R)^t(\s\eta) (\s R)=\s\eta$, where
$\s\eta=S\eta S=-\eta$.  
This proves that $\s$ is a group automorphism for $SO(3,3)$.
Notice that $\s^2={\rm id}$. Since $S$ is not an element of
$SO(3,3)$, this automorphism is called external.
The $\s$ automorphism has naturally a counter part
which acts on the $so(3,3)$ algebra.
The $\s$-invariant
sub-algebra can be proved to be
isomorphic to $gl(3)$. Since $gl(3)$ does not contain any $so(1,3)$
subalgebra, there is no $\s$-invariant $so(1,3)$ subalgebra in $so(3,3)$
and, as groups are concerned, there is neither any $\s$-invariant
embedding of $SO(1,3)$ in $SO(3,3)$.

$so(1,3)$ subalgebras of $so(3,3)$
are specified as commutants with non invariant $so(2)$
subalgebras. There exists two inequivalent classes of them
up to group conjugation. The two relative Cartan generators get 
in fact exchanged by $\s$ as $\s t_\pm=t_\mp$ (up to group conjugation).
Notice that, given one of such $so(1,3)$
subalgebras one can slightly extend it to a one parameter family of 
$iso(1,3)$ algebras,
parametrized by a $SO(2)$ angle, relative to the possible projection 
choices of half of the off block diagonal elements with respect to the
$SO(2)\times SO(1,3)$ index decompositions.

\end{document}